\documentclass[letterpaper,twocolumn,10pt]{article}
\usepackage{usenix-2020-09}

\RequirePackage{caption}
\usepackage{paralist,tabularx}
\usepackage{todonotes}
\usepackage{ctable}
\usepackage{hyperref}
\usepackage{subfig}
\usepackage{amssymb,amsmath}
\usepackage{graphicx}
\usepackage{balance}
\usepackage{lscape}
\usepackage{multirow}
\usepackage{xurl}
\usepackage{xcolor}
\usepackage{colortbl}
\usepackage{algorithm2e}
\usepackage{listings}
\usepackage{authblk}
\definecolor{codegreen}{rgb}{0,0.6,0}
\definecolor{codegray}{rgb}{0.5,0.5,0.5}
\definecolor{codepurple}{rgb}{0.58,0,0.82}
\definecolor{backcolour}{rgb}{0.95,0.95,0.92}

\lstdefinestyle{mystyle}{
  backgroundcolor=\color{backcolour}, commentstyle=\color{codegreen},
  keywordstyle=\color{magenta},
  numberstyle=\tiny\color{codegray},
  stringstyle=\color{codepurple},
  basicstyle=\ttfamily\footnotesize,
  breakatwhitespace=false,         
  breaklines=true,                 
  captionpos=b,                    
  keepspaces=true,                 
  numbers=left,                    
  numbersep=5pt,                  
  showspaces=false,                
  showstringspaces=false,
  showtabs=false,                  
  tabsize=2
}

\newif\ifremarks

\remarksfalse 
\remarkstrue 

\begin{document}

\title{Beyond the Surface: Investigating Malicious CVE Proof of Concept Exploits on GitHub}

\author{Soufian El Yadmani, Robin The, Olga Gadyatskaya}
\affil{Leiden Institute of Advanced Computer Science, Leiden University}

\maketitle

\begin{abstract}
\

Exploit proof-of-concepts (PoCs) for known vulnerabilities are widely shared in the security community. They help security analysts to learn from each other and they facilitate security assessments and red teaming tasks. In the recent years, PoCs have been widely distributed, e.g., via dedicated websites and platforms, and public code repositories such as GitHub. However, there is no guarantee that PoCs in public code repositories come from trustworthy sources or even that they do what they are supposed to do.

In this work we investigate GitHub-hosted PoCs for known vulnerabilities discovered in 2017--2021. We discovered that not all PoCs are trustworthy. Some proof-of-concepts are malicious, e.g., they attempt to exfiltrate data from the system they are being run on, or they try to install malware on this system, and in some cases they have hard-coded reverse shell listener. 

To measure the prevalence of this threat, we propose an approach to detecting malicious PoCs. Our approach relies on the maliciousness symptoms we have observed in our PoC dataset: calls to malicious IP addresses, encoded malicious code, and included Trojanized binaries. With this approach, we have discovered 899 malicious repositories out of 47,285 repositories that have been downloaded and checked (i.e., 1.9\% of the studied repositories have indicators of malicious intent). This figure shows a worrying prevalence of dangerous malicious PoCs among the exploit code distributed on GitHub.

\end{abstract}

\section{Introduction}\label{sec:intro}

CVE, which stands for \emph{Common Vulnerabilities and Exposures}\footnote{\url{https://www.cve.org/}}, is a comprehensive list of publicly disclosed security flaws in software or systems. These flaws are identified by a unique CVE ID. During penetration testing or security assessments, pentesters strive to identify these known vulnerabilities in customers' environments so that they can be patched. However, it is not enough to simply locate a vulnerable system; pentesters must also demonstrate its \emph{exploitability}. Professional frameworks like Metasploit\footnote{\url{https://www.metasploit.com/}} and reputable databases such as Exploit-DB\footnote{\url{https://www.exploit-db.com/}} offer exploits for many CVEs. However, not all CVEs have corresponding exploits publicly available~\cite{jacobs2021exploit,householder2020historical,nayak2014some,figueiredo2023statistical}. To find potential \emph{Proof of Concept} (PoC) exploits that demonstrate vulnerabilities, pentesters then often turn to public code repositories such as GitHub\footnote{\url{https://github.com/}}.

While reputable sources such as Exploit-DB validate the effectiveness and legitimacy of PoCs, public code platforms such as GitHub lack an exploit vetting process. Pentesters may consider certain repository properties including but not limited to the number of watchers, stars, and forks to gauge the popularity and potential utility of a specific PoC. However, the absence of vetting on GitHub means that PoCs found there may be unreliable or even contain malicious content, posing a risk to the person running them, especially if on a customer's infrastructure.

The security community has previously discussed the lack of trustworthiness of PoCs published on GitHub and social media platforms. For example, reputable cyber security blogs like Bleepingcomputer\footnote{\url{https://www.bleepingcomputer.com/news/security/fake-windows-exploits-target-infosec-community-with-cobalt-strike/}} have reported instances of PoCs on GitHub containing the CobaltStrike\footnote{\url{https://www.cobaltstrike.com/}} backdoor, which targeted the security community with a fake exploit for \texttt{CVE-2022-26809}\footnote{\url{https://nvd.nist.gov/vuln/detail/CVE-2022-26809}}. Security researcher Curtis Brazzell conducted investigations by setting up honey-PoCs and found a remarkably high number of people running unverified PoCs from GitHub\footnote{\url{https://curtbraz.medium.com/exploiting-the-exploiters-46fd0d620fd8}}.

In our research, we conduct a comprehensive investigation into the distribution of \emph{malicious PoCs} on GitHub. To accomplish this, we collected publicly available PoCs shared on GitHub for CVEs discovered between 2017 and 2021. Our dataset comprises 47,285 repositories sharing PoCs for at least one CVE within the specified time frame. To the best of our knowledge, this is the first large-scale empirical investigation of malicious PoCs. We developed a set of heuristics that identify malicious indicators within PoCs for detecting potentially harmful exploits. Applying these heuristics, we discovered 899 malicious repositories, which accounts for 1.9\% of the total PoC repositories in our dataset.

Additionally, we examined GitHub repository meta data, such as the number of forks and stars for both malicious and non-malicious PoC repositories, as well as code similarity between malicious and non-malicious repositories. Our analysis revealed that malicious repositories exhibit higher similarity to each other compared to non-malicious repositories. Therefore, code similarity metrics can aid in identifying new malicious PoCs. However, our findings indicate that the numbers of forks and stars are not effective in discriminating between malicious and non-malicious PoCs.


As PoCs are widely used for learning and security assessments, the presence of malicious PoCs poses significant security risks. Our research sheds light on this issue by proposing an approach to identify symptoms of malicious intent in PoCs. The discovery of 1.9\% of the studied repositories exhibiting such symptoms showcases the alarming prevalence of dangerous PoCs on GitHub. By uncovering this threat and highlighting the need for caution and scrutiny when utilizing PoCs, our work contributes to the urgent task of enhancing the trustworthiness and reliability of shared security resources in the community.

To facilitate further community investigations into PoCs, including malicious ones, we share our dataset and provide an implementation of our approach\footnote{More information on how to access these resources will be included in the final print due to the anonymization requirements.}.

\paragraph{Responsible Disclosure.} 
 We have reported our findings to GutHub via their dedicated responsible disclosure channel in October 2022.

\section{Data Collection and Analysis}\label{sec:data}

\begin{figure*}[t!]
    \centering
    \captionsetup{justification=centering}
    \includegraphics[scale=0.35]{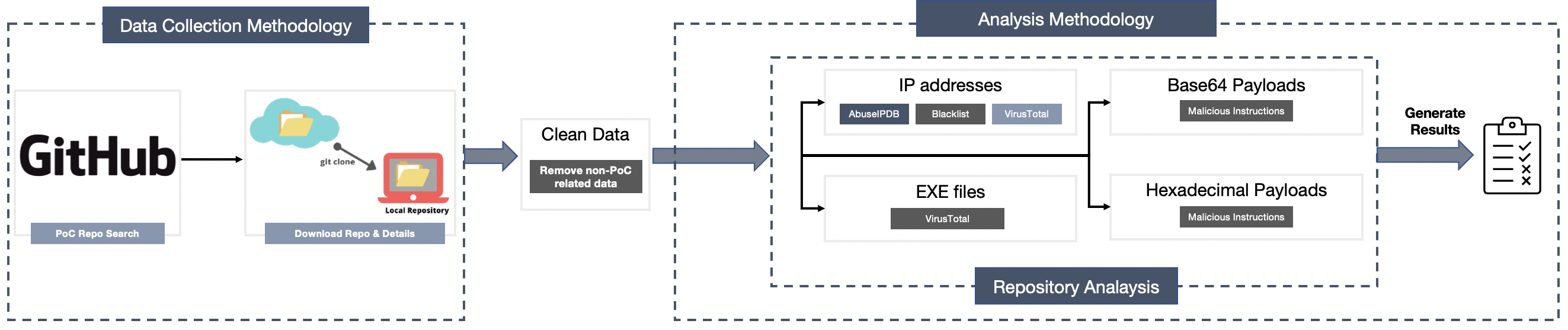}
  \caption{Data collection and analysis methodology}
  \label{fig:AnalysisAndCollection}
\end{figure*}\

Figure~\ref{fig:AnalysisAndCollection} shows the methodology of our study. To achieve our goal of analyzing all PoCs on GitHub made for specific CVEs, we first gather a dataset, then clean it and process it.

\subsection{Data Collection}\label{sec:datacollection}



\begin{figure}[t]
    \centering
    \captionsetup{justification=centering}
    \includegraphics[scale=0.32]{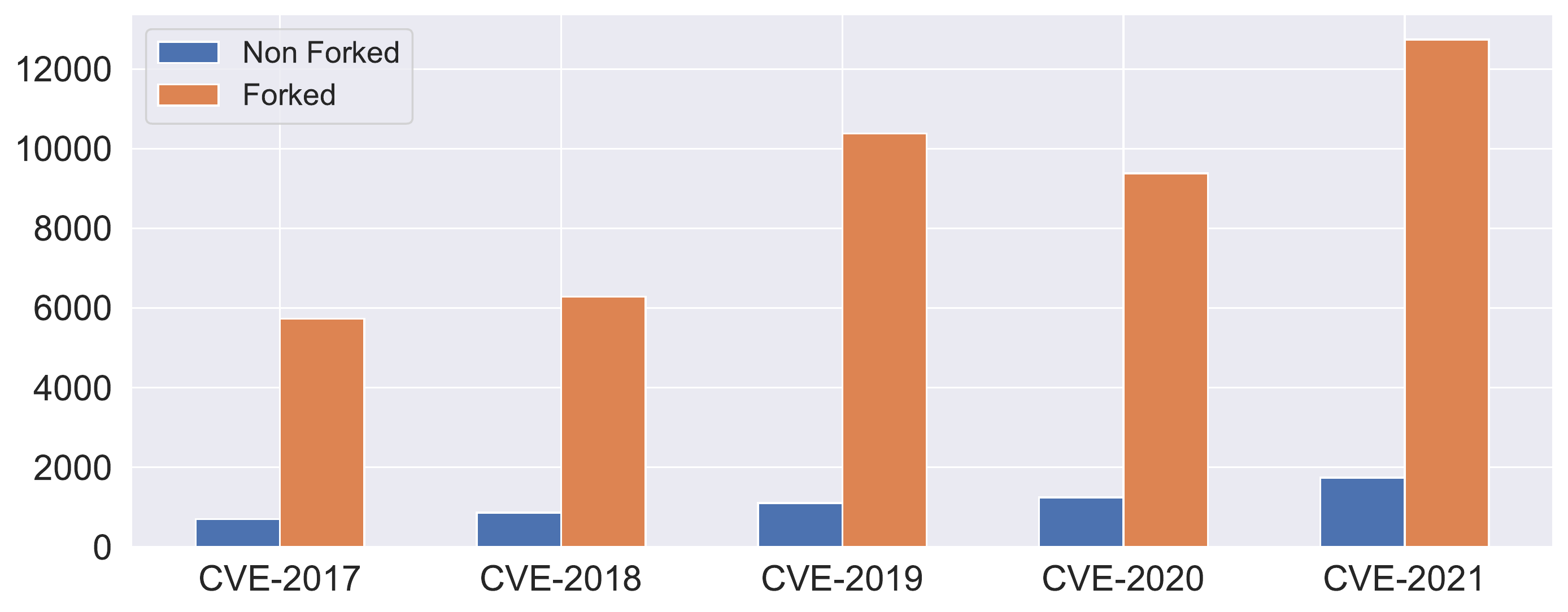}
  \caption{Downloaded GitHub repositories containing PoCs for CVEs}
  \label{fig:over_down}
\end{figure}

To gather the data, we used the GitHub API, which provides keyword-based search capabilities for repositories, code, and commits. Our focus was on finding repositories that contained Proof of Concepts (PoCs) for CVEs. We have also collected meta data about these repositories, including  descriptions, star ratings, and fork counts. Our dataset spans five full years, from 2017 to 2021, encompassing CVEs assigned during this period (i.e., CVE IDs with prefixes \texttt{CVE-2017}, \texttt{CVE-2018}, \texttt{CVE-2019}, \texttt{CVE-2020} and \texttt{CVE-2021}).

Initially, we compiled a list of CVE IDs issued within our target period from MITRE\footnote{\url{https://cve.mitre.org/cve/}}. We collected data by searching repositories for keywords following the specific CVE IDs in the format ``\texttt{CVE-YEAR-ID}''\footnote{\url{https://cve.mitre.org/cve/identifiers/syntaxchange.html}}. We also used variations of keywords, such as ``\texttt{CVE YEAR-ID}'', ``\texttt{CVE-YEAR ID}'', ``\texttt{CVE YEAR ID}'' and ``\texttt{CVE:YEAR-ID}'' to increase our search coverage. According to the GitHub API documentation\footnote{\url{https://docs.github.com/en/rest/search\#search-repositories}}, the keyword search is performed on the repository name and description. We then have analyzed the collected repositories for mentions of CVE IDs within their README file and code files to map the collected repositories to the targeted CVEs in a more precise way.

We collected all available repositories that mentioned CVEs discovered within our target period, including forked repositories. The data was downloaded and stored between April 10, 2022, and April 23, 2022. 

\subsection{Data Cleaning}

After our data collection process, we then cleaned the dataset. As the GitHub search is keyword-based, the results obtained were the outcome of searching within the repository names and descriptions. As a consequence, the search results contained various repositories that were not all specifically PoCs, but rather Indicators of Compromise (IoCs) and descriptions of CVEs.
To ensure the accuracy and relevance of the dataset, we performed additional filtering focusing on reviewing the repositories that contained many CVE IDs or included explicit IoC strings (``indicator of compromise'' or ``IoC'').  We examined the contents, source code, and associated documentation of such repositories to verify if they contained PoCs for CVE exploits. This cleaning process enabled us to eliminate irrelevant repositories and focus on those that aligned with our research objectives. As a result of this cleaning procedure, the number of repositories under analysis was refined from 48,700 to 47,285, entailing the exclusion of 1,415 repositories.

Figure~\ref{fig:over_down} depicts the number of repositories related to CVEs after cleaning, including both original and forked repositories. As illustrated in the figure, the number of repositories containing PoCs has steadily increased over the years.   

\subsection{Data Analysis}

After cleaning the data, we first analyzed the prevalence of programming languages during the specified period based on the language labels assigned by GitHub. The statistics regarding programming languages are presented in Table~\ref{table:prog_lang}. It is evident from the table that Python has emerged as the dominant programming language among hackers and exploit developers over the past five years. This can be attributed to the extensive range of libraries available in Python that support fundamental programming and ``hacking'' tasks. It is worth noting that GitHub was unable to label the language of the majority of repositories due to their inclusion of various file types, programming languages, or solely consisting of README files that describe the attack. Consequently, these repositories were categorized as ``Undetected'' in our dataset.

\begin{table}[t]
\centering
\caption{Overview of top 10 used programming languages}
\label{table:prog_lang}
\arrayrulecolor[rgb]{0.541,0.541,0.541}
\begin{tabular}{ll} 
\arrayrulecolor{black}\hline
\textbf{Programming language}                         & \textbf{Count}  \\ 
\arrayrulecolor[rgb]{0.541,0.541,0.541}\hline
HTML                                         & 296    \\
\rowcolor[rgb]{0.804,0.804,0.804} Ruby       & 379    \\
Go                                           & 400    \\
\rowcolor[rgb]{0.804,0.804,0.804} JavaScript & 548    \\
Shell                                        & 652    \\
\rowcolor[rgb]{0.804,0.804,0.804} C++        & 962    \\
Java                                         & 1071   \\
\rowcolor[rgb]{0.804,0.804,0.804} C          & 1686   \\
Python                                       & 8305   \\
\rowcolor[rgb]{0.804,0.804,0.804} 
Undetected & 31830 \\ \hline
\end{tabular}
\arrayrulecolor{black}

\end{table}

For a better understanding of our PoCs dataset and targeted issues, we conducted a cross-check with the NIST NVD database\footnote{\url{https://nvd.nist.gov/}} to determine the number of CVEs for which PoCs were available on GitHub. The results of this analysis are summarized in Table~\ref{table:CVEs_unique}. It is evident from the table that only a small portion of CVEs have corresponding public exploit code published on GitHub. The number of PoCs listed in Table~\ref{table:CVEs_unique} is higher than the total number of repositories collected, as many repositories contain PoCs for multiple CVEs, sometimes spanning different years.

\begin{table*}[t]
\centering
\caption{Overview of the collected data with respect to unique CVE IDs and number of repositories and PoCs}
\label{table:CVEs_unique}
\begin{tabular}{ccccc} 
\hline
\textbf{CVE-Year}                                   & \textbf{\# Unique CVEs targeted} & \% \textbf{CVEs assigned by NVD} & \textbf{\# Repos} & \textbf{\# PoCs} \\ 
\hline
2017                                   & 338                     & 2.30\%                  & 6424    & 7731   \\
\rowcolor[rgb]{0.804,0.804,0.804} 2018 & 447                     & 2.70\%                  & 7021    &  8930  \\
2019                                   & 606                     & 3.49\%                  & 11336   & 14321 \\
\rowcolor[rgb]{0.804,0.804,0.804} 2020 & 726                     & 3.96\%                  & 10588   & 23255\\
2021                                   & 658                     & 3.26\%                  & 14487   & 17607\\ 
\hline
Total                           & 2775                    & 3.19\%                  & 47285 (unique)   & 72580
\end{tabular}

\end{table*}



\begin{figure}[t]
    \centering
    \captionsetup{justification=centering}
    \includegraphics[scale=0.28]{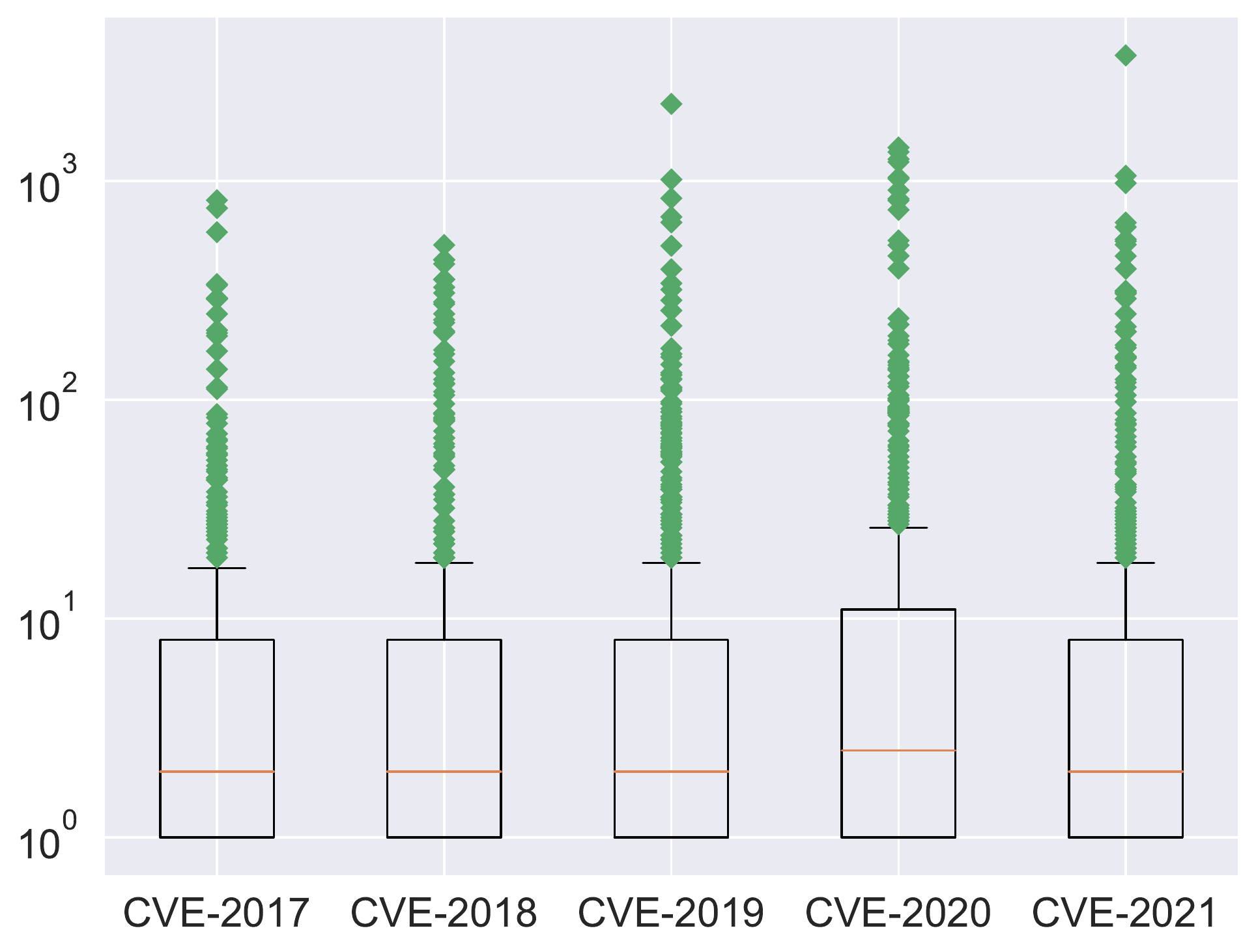}
    \caption{Distribution of CVE PoCs per year}
  \label{fig:CVEsPerYear}
\end{figure}

By analyzing the PoCs data, we can determine which CVEs have the highest number of PoCs available on GitHub. Figure~\ref{fig:CVEsPerYear} presents a boxplot illustrating the distribution of PoCs for CVEs on GitHub during the target period. Each data point on the graph represents a unique CVE ID and the corresponding number of PoCs found in our dataset.

Overall, the number of PoCs per CVE on GitHub has remained relatively consistent across the years. This indicates that the availability of PoCs on GitHub is not heavily influenced by the year of CVE discovery.

The three points at the top of Figure~\ref{fig:CVEsPerYear}, representing the CVEs with the highest number of PoCs, correspond to arguably the most significant security issues of the past decade. The foremost outlier is \texttt{CVE-2021-44228}\footnote{\url{https://nvd.nist.gov/vuln/detail/CVE-2021-44228}} (also known as Log4Shell), a critical vulnerability in Log4j. The second CVE with the most PoCs is \texttt{CVE-2019-0708}\footnote{\url{https://nvd.nist.gov/vuln/detail/CVE-2019-0708}} (also known as BlueKeep), which pertains to a vulnerability in the Remote Desktop Protocol (RDP). \texttt{CVE-2020-1938}\footnote{\url{https://nvd.nist.gov/vuln/detail/CVE-2020-1938}}, a vulnerability in the Apache JServ Protocol, is in the third position. It is noteworthy that some of these flaws continue to be widely exploited by hackers in the wild\footnote{\url{https://blog.checkpoint.com/security/april-2023s-most-wanted-malware-qbot-launches-substantial-malspam-campaign-and-mirai-makes-its-return/}}, explaining their prominent positions in our dataset in terms of the number of published PoCs.

Table~\ref{table:repos_by_securityissue} presents the distribution of CVE types in our dataset of PoCs collected from GitHub. The table categorizes the CVEs based on different security issues associated with them. Each row represents a specific security issue type, and the corresponding number of CVEs related to that issue, which was collected from CVEDetails\footnote{\url{https://www.cvedetails.com/}} for each CVE according to its CWE (Common Weakness Enumeration). The table offers an overview of the types of security issues that are prevalent in the dataset. We can see that the collected PoCs target a large variety of security issues, with code execution being the most prevalent weakness type.

\begin{table}[t]
\centering
\captionsetup{justification=centering}
\caption{Distribution of CVEs in our data by issue according to \url{https://www.cvedetails.com/}}
\label{table:repos_by_securityissue}
\begin{tabular}{ll}
\hline
\textbf{Security issue}         & \textbf{\# of CVEs} \\ \hline
Overflow                        & 184                 \\
\rowcolor[rgb]{0.804,0.804,0.804} 
Gain privileges                 & 47                  \\
Denial Of Service               & 176                 \\
\rowcolor[rgb]{0.804,0.804,0.804} 
Execute code                    & 743                \\
Obtain information              & 108                 \\
\rowcolor[rgb]{0.804,0.804,0.804} 
Memory corruption               & 76                 \\
Bypass a restriction or similar & 169                 \\
\rowcolor[rgb]{0.804,0.804,0.804} 
Cross-site scripting            & 272                 \\
SQL injection                   & 97                  \\
\rowcolor[rgb]{0.804,0.804,0.804} 
Directory traversal             & 129                 \\
CSRF                            & 51                  \\
\rowcolor[rgb]{0.804,0.804,0.804} 
File inclusion                  & 14                  \\
Untagged                        & 1092                \\
\hline
\end{tabular}

\end{table}


\begin{figure}[t]
    \centering
    \captionsetup{justification=centering}
    \includegraphics[scale=0.34]{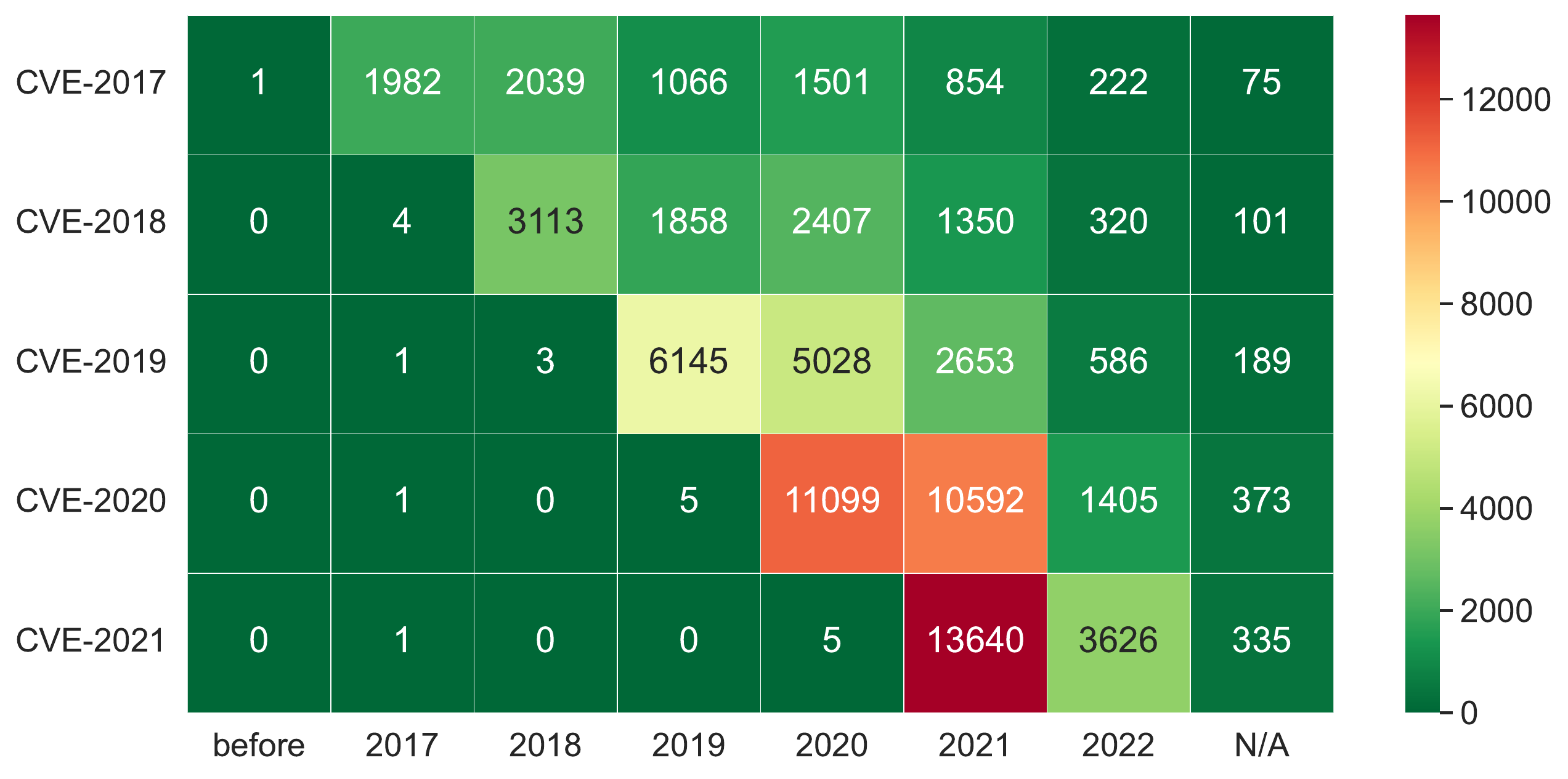}
    \caption{Heatmap of the repositories per CVE-year and the year of creation of the repository}
  \label{fig:CVEsdistributionPoCsCreate}
\end{figure}

\begin{figure}[t]
    \centering
    \captionsetup{justification=centering}
    \includegraphics[scale=0.34]{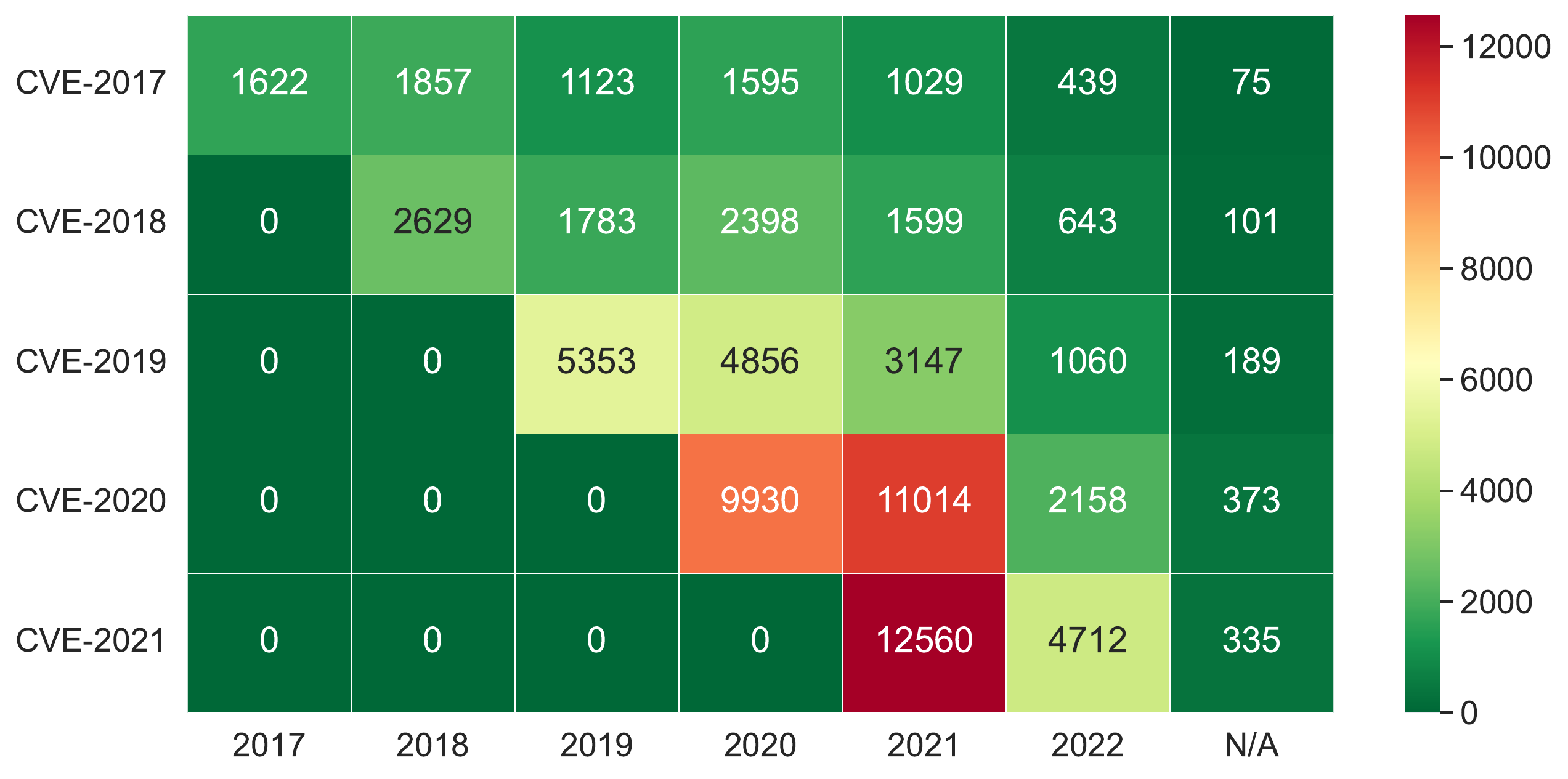}
    \caption{Heatmap of the repositories per CVE-year and the year of last update to the repository}
  \label{fig:CVEsdistributionPoCs}
\end{figure}

To illustrate the relationship between the targeted CVE-year and the year of repository creation, we present a heatmap in Figure~\ref{fig:CVEsdistributionPoCsCreate}. It is evident that the majority of repositories were created in the same year as the corresponding CVE. However, there are also instances where repositories were created prior to the issuance of the associated CVE. This is attributed to repositories containing PoCs for multiple CVEs (we verified these cases manually).

Furthermore, Figure~\ref{fig:CVEsdistributionPoCs} illustrates the distribution of repositories based on the targeted CVE-year and the year of the last update. Both figures indicate that the highest number of PoC repositories were created and last updated in 2021. Repositories for older CVEs continue to be created and updated even several years later.

\section{Malicious PoCs}\label{sec:fakepocs}

In this section, we present our proposed method for detecting malicious proof of concepts (PoCs) in GitHub. We begin by discussing our definition of malicious PoCs and observations on indicators for detecting these PoCs. Figure~\ref{fig:AnalysisAndCollection} provides an overview of the steps undertaken during our research to analyze the PoC repositories. 


\paragraph{Our threat model.}
We first define what kind of PoCs we are looking for (what do we consider to be \emph{malicious PoCs}). Typically, PoCs are developed by hackers or security professionals with the intention of exploiting vulnerabilities in specific products to compromise systems. Consequently, all PoCs inherently represent malicious software that would trigger security systems designed to detect exploits. Moreover, they may exhibit additional properties associated with malicious software, such as inclusion of malicious binaries, e.g., hacking tools. 

As \emph{malicious PoCs} we consider PoCs that include adversarial functionality beyond the code necessary for exploiting the targeted CVE. A typical example would be a PoC that, when executed, attacks not the system specified by the security professional as the target, but the system of the security professional. In Section~\ref{sec:casestudies} we discuss concrete examples of such behavior in more details.  

One of the primary challenges we encountered in this study was identifying reliable indicators of a malicious PoC.  Therefore, our task was to identify characteristics of such software that are unexpected in legitimate PoCs intended only for exploiting the targeted system, but instead indicate adversarial goals unrelated to the original PoC objectives. In the remainder of this section we discuss the heuristics we propose for identifying such adversarial functionality and present the detection results.

\subsection{Proposed Heuristics}\label{sec:Reasoning}

We have identified the following indicators of malicious PoCs.



\begin{itemize}
\item \textbf{IP Indicators}: In general, PoCs should not have communications with predefined public IP addresses, except for downloading required tools. Communications with IP addresses could indicate malicious behavior, such as exfiltrating information from the executing machine or/and download malicious files. To analyze this, we extracted all IP addresses from the repositories using regular expressions. We then filtered for public IP addresses, removing private IP addresses and IPs mentioned in comments, help menus, and IoCs. This reduced the number of IP addresses we needed to check manually. Next, we compared the extracted IP addresses with public blacklists and cross-referenced information on VirusTotal and AbuseIPDB.

\item \textbf{Binary analysis}: Some PoC repositories provide pre-built binaries to simplify the exploitation process. We focused on EXE files that can run on Windows systems, as most malware attacks target Windows users\footnote{\url{https://www.neowin.net/news/2022-sees-over-5000-times-new-windows-malware-vs-macos-over-60-times-vs-linux/}}. After extracting the binaries, we checked their hashes on VirusTotal to determine their safety. However, analyzing hacking tools on VirusTotal requires caution due to detection and behavioral considerations. We checked if the file was flagged as malicious by VirusTotal and then examined whether the binary was flagged as an exploit for the target CVE or as a known hacking tool. This approach helped us exclude this type of binaries, and only focus on the rest which are flagged for other reasons. We note that binary analysis with VirusTotal is a common way to identify malicious functionality~\cite{al2022no}.

\item \textbf{Analysis of obfuscated payloads}: Encoding is frequently used by adversaries to hinder analysis, and extracting encoded payload provides insights into potential malicious instructions or hidden commands within the code~\cite{ohm2022feasibility,ladisa2022towards}.

\begin{itemize}
\item \textbf{Hexadecimal analysis}: Hexadecimal encoding is commonly used to obfuscate malicious payloads. We extracted hexadecimal payloads from all code files and concatenated the values for each file. When possible, we decoded the concatenated values into a human-readable format to identify IP calls or other payloads that might be encoded within.
\item \textbf{Base64 analysis}: Base64 encoding is another prevalent method for obfuscating malicious payloads. By extracting base64 values using regular expressions, we analyzed them and decoded hidden scripts to detect any connections with external IPs within the encoded payloads.

\end{itemize}

\end{itemize}

These techniques collectively enhance our ability to flag and mitigate the risks associated with malicious PoCs, helping to protect users within the security community and raise awareness of potential threats. In the following sections, we delve into the details of each analysis and present the outcomes of our study.
\subsection{IP Analysis}\label{sec:IP}

To extract public IP addresses from the code, we employed a regular expression pattern: (\texttt{(25[0-5]|2[0-4][0-9]|[01]?[0-9][0-9]?) (\textbackslash.(25[0-5]|2[0-4][0-9]|[01]?[0-9][0-9]?))\{3,\}}). This pattern allowed us to identify IP addresses within the source code. Our filtering process aimed to include only public IP addresses that could potentially function as Command \& Control servers or be involved in downloading malicious files. To achieve this, we excluded private IP addresses based on the guidelines outlined in \texttt{RFC1918}\footnote{\url{https://datatracker.ietf.org/doc/html/rfc1918}}. Private IP addresses  are reserved for internal use within private networks and are not intended to be publicly accessible. 

It is important to note that the identified IP addresses could also appear in help menus, comments, or examples within the code. Automatically detecting how these IPs are used within the code proved challenging. To mitigate this limitation, we manually reviewed PoC repositories and files triggered by the previous regular expression, to assess how these IPs were being used in the code. After applying the relevant IP filtering and establishing a subset of public IP addresses actually used in PoCs, we performed a comprehensive analysis of the extracted IPs. Our analysis consisted of multiple steps to ensure a thorough assessment of their malice.




Firstly, we cross-referenced the filtered IP list with the blacklists (blocklists) provided by the FireHOL\footnote{\url{https://iplists.firehol.org/}} IP Lists project. These blacklists, available on GitHub\footnote{\url{https://github.com/firehol/blocklist-ipsets}} and updated daily, contain a wide range of IPs associated with malware and ransomware campaigns, Command\&Control (C\&C) servers, and compromised machines. By comparing our IP list with these blacklists, we could identify any IPs that have already been flagged due to their malicious activities.

The next step involved checking these IPs against VirusTotal. We aimed to determine if any of the IPs had been previously identified as malicious by security vendors. Additionally, we performed scans of these IPs using VirusTotal's comprehensive scanning capabilities. This approach allowed us to gather detailed information about the IPs, such as their reputation, associated malware, and any available reports or detections.

If an IP was not flagged as malicious and is currently inactive, it can be challenging to ascertain its past malicious activities. Unless the IP's malicious behavior was captured and shared by other researchers through social media or public reports, its prior malice remains difficult to determine with certainty. Thus, to further enhance our IP analysis, we leveraged AbuseIPDB\footnote{\url{https://www.abuseipdb.com/}}, which relies on historical reports of malicious IP activities. AbuseIPDB provides valuable insights, including the number of times an IP has been reported, the date of the reports, and the types of attacks associated with the IP. By cross-referencing our IP list with AbuseIPDB, we gained an additional layer of analysis to validate and corroborate the maliciousness of the IPs.

The results of our comprehensive IP analysis, including the findings from the FireHOL IP Lists cross-matching, VirusTotal scans, and AbuseIPDB checks, are presented in detail in Table~\ref{table:ipscan}. These results provide insight into the reputation and potential malicious activities associated with the extracted IPs. The number of malicious IPs over all repositories is not high compared to the number of repositories analyzed. Still, we think that this heuristic is useful indicator for pentesters to spot malicious PoCs. 

\begin{table*}[t!]
\centering
\captionsetup{justification=centering}
\caption{Collection, validation and detection results of malicious IPs through blacklists, VirusTotal, and AbuseIPDB}
\label{table:ipscan}
\arrayrulecolor{black}
\begin{tabular}{cccccc}
\hline
\textbf{Year} &  \textbf{Extracted IPs patterns} & \textbf{Blacklist} & \textbf{VirusTotal} & \textbf{Abuse IPDB} & \textbf{Total malicious IPs} \\ \hline
2017  & 203498        & 0         & 20         & 1          & 21                  \\  \rowcolor[rgb]{0.804,0.804,0.804}
2018  & 10252         & 0         & 0          & 0          & 0                   \\
2019  & 268120        & 1         & 1          & 0          & 1                   \\  \rowcolor[rgb]{0.804,0.804,0.804}
2020  & 107153        & 0         & 0          & 0          & 0                   \\
2021  & 88118        & 0         & 0          & 0          & 0                   \\ \hline
Total & 677141        & 1         & 21         & 1          & 22                  \\ 
\end{tabular}

\end{table*}

\subsection{Binary Analysis}\label{sec:binary}

To gain insights into the nature and potential maliciousness of the EXE binaries included within the PoCs, we conducted a comprehensive analysis using VirusTotal. VirusTotal is a widely trusted platform often employed by researchers for this purpose~\cite{de2021compromised,shen2022large,graziano2015needles,al2022no,zhu2020measuring}. 

Our analysis began by collecting the unique \texttt{sha256} hash of each EXE binary encountered across all repositories. VirusTotal maintains an extensive database of previously scanned binaries, indexed by their respective hash values. By querying this database with the collected hashes, we were able to access existing reports associated with the majority of the binaries. These reports provided valuable insights into potential malware presence or any other malicious activities associated with the binaries.

For the hashes that were not present in VirusTotal's database, we took additional measures to ensure a comprehensive analysis. We uploaded these binaries to VirusTotal for scanning, which generated detailed reports outlining their characteristics and flagged any potential malicious behavior.

Our approach relies on leveraging the capabilities of VirusTotal to identify and flag malicious binaries that exploit specific CVEs. This feature provided valuable insights into distinguishing between binaries that serve as exploits for targeted vulnerabilities, binaries that are known to be hacking tools\footnote{The excluded tools contained many versions and variants of \texttt{Netcat}\footnotemark, \texttt{Rubeus}\footnotemark, \texttt{PowerSploit}\footnotemark, \texttt{Ysoserial}\footnotemark, and other known tools.}, and those with malicious intent unrelated to the repository's intended use.  The algorithmic process guiding our decision-making on binary maliciousness in our context is outlined in Algorithm~\ref{alg:AlgoVT}. The results of this approach are summarized in Table~\ref{table:binaries_hacktools_cves}. 

\addtocounter{footnote}{-3} 
\footnotetext{\url{https://netcat.sourceforge.net/}}
\stepcounter{footnote}\footnotetext{\url{https://github.com/GhostPack/Rubeus}}
\stepcounter{footnote}\footnotetext{\url{https://github.com/PowerShellMafia/PowerSploit}}
\stepcounter{footnote}\footnotetext{\url{https://github.com/frohoff/ysoserial}}

\begin{algorithm}[t]

\SetAlgoLined 
\KwData{Hashes of all extracted EXE binaries} 
\KwResult{\texttt{True} if malicious, else \texttt{False}} 
\ForEach(){Hash}{
    \If{Hash of binary is not known by VirusTotal}{
        Upload the binary file to VirusTotal\;
    }
    \eIf{Binary is marked malicious by VirusTotal} {
        \eIf{Maliciousness is related to the CVE-ID in the studied repository \textbf{OR} Maliciousness is related to a known hacking tool}{
            return \texttt{False};
        }{
            return \texttt{True};
        }
    }{
        return \texttt{False};
    }
}
\captionsetup{justification=centering}
\caption{Procedure to check if binary is malicious on VirusTotal} 
\label{alg:AlgoVT}
\end{algorithm}


\begin{table}[t]
\centering
\captionsetup{justification=centering}
\caption{Binary labels based on VirusTotal scan}
\label{table:binaries_hacktools_cves}
\arrayrulecolor{black}
\begin{tabular}{cc}
\arrayrulecolor{black}\hline
\textbf{Label}  &  \textbf{\# Unique binaries}\\
\arrayrulecolor[rgb]{0.541,0.541,0.541}\hline
Malicious  &     136       \\
\rowcolor[rgb]{0.804,0.804,0.804} CVE Related  &     19        \\
Hacktool  &     29        \\ \arrayrulecolor{black}\hline
Total  &     284        \\
\end{tabular}

\end{table}

Through this rigorous analysis using VirusTotal, we were able to gain valuable insights into the nature of the EXE binaries included in the PoCs. These findings played a crucial role in our overall investigation of potential malicious activities and helped in determining the level of risk associated with each binary.

The results of our binary analysis are summarized in Table~\ref{table:malbinaries}. Notably, we observed a significant number of repositories containing EXE binaries, indicating that attackers often use repositories to distribute malware that can deceive unsuspecting users. Furthermore, the presence of a substantial number of malicious binaries, particularly between 2019 and 2021, highlights the prevalence of such threats. It is worth mentioning that we also encountered repositories containing multiple malicious binaries, as indicated in the table, emphasizing the varied strategies employed by attackers to propagate their malicious software.

\begin{table}[h]
\centering
\captionsetup{justification=centering}
\caption{Repositories with malicious binaries }
\label{table:malbinaries}
\arrayrulecolor{black}
\begin{tabular}{cccc}
\hline
\textbf{Year}  &  \textbf{\# Binaries} &   \textbf{\# Malicious}  & \textbf{\# Malicious}     \\
& & \textbf{binaries} & \textbf{repos} \\
\arrayrulecolor[rgb]{0.541,0.541,0.541}\hline
2017  &     423        & 126   & 99      \\
\rowcolor[rgb]{0.804,0.804,0.804} 2018  &     1269        & 52   & 33     \\
2019  &     1678       & 537  & 255       \\
\rowcolor[rgb]{0.804,0.804,0.804} 2020  &     2068       & 380    &174     \\
2021  &     804        & 274  & 227 \\ \arrayrulecolor{black}\hline
Total  &     6160       & 1353    &782     \\
\end{tabular}

\end{table}

\subsection{Hexadecimal Analysis}\label{sec:Hex}

PoCs often include hexadecimal payloads as part of the exploit. However, also attackers frequently employ hexadecimal encoding as an obfuscation technique, making it a prime focus of our research. As a result, we extracted encoded hexadecimal payloads from the PoCs using a regular expression pattern, specifically \texttt{x[A-Fa-f0-9]+}, and subsequently decoded them for the purpose of analyzing potential malicious code. The number of PoCs containing hexadecimal code over the past five years is presented in Table~\ref{table:hexresults}.

\begin{table}[h]
\centering
\captionsetup{justification=centering}
\caption{Results of hexadecimal scan}
\label{table:hexresults}
\arrayrulecolor{black}
\begin{tabular}{cccc} 
\hline
\textbf{Year}  & \textbf{\# Repos}    & \textbf{\# Containing hex} & \textbf{\# Malicious hex}            \\ 
\arrayrulecolor[rgb]{0.541,0.541,0.541}\hline
2017                                   & 6424   & 3367       & 0                      \\
\rowcolor[rgb]{0.804,0.804,0.804} 2018 & 7021    & 3511        & 0                      \\
2019                                   & 11336 & 7377     & 0                     \\
\rowcolor[rgb]{0.804,0.804,0.804} 2020 & 10588   & 7140       & 41                     \\
2021                                   & 14487   & 8281       & 0 \\ \arrayrulecolor{black}\hline
Total  &     47285       & 29676    &41     \\
\end{tabular}

\end{table}

Our observations indicate a steady increase in the occurrences of hexadecimal code over the past five years. After decoding all hexadecimal code strings and categorizing them as suspicious or not (based on the inclusion of a public IP address), we discovered 41 repositories utilizing hexadecimal code as a means to obfuscate malicious code, thereby evading detection by readers. These payloads were decoded and executed to carry out various types of instructions. It is noteworthy that all the malicious payloads we encountered were from the year 2020 and were linked to a study conducted by security researcher Curtis Brazzell. He established honey-PoCs and measured the number of individuals who executed them without inspecting the code\footnote{\url{https://curtbraz.medium.com/exploiting-the-exploiters-46fd0d620fd8}}.

\subsection{Base64 Analysis}\label{sec:Base64}

Base64 is an encoding system commonly employed to embed binary assets within textual assets, such as programming languages. Proof of concept authors used this encoding system to incorporate payloads that facilitate the exploitation of vulnerabilities associated with specific CVEs. Due to the widespread use of this technique in exploits, attackers have also adopted it to embed additional code within their malicious proof of concepts. This embedded code is subsequently executed by the script, following the provided instructions. Some of these instructions involve exfiltrating information, while others involve downloading and executing malicious files on the system.
During our analysis, we initially conducted an automated search for base64 payloads using the following regular expression in our proof of concepts:\\
(\texttt{[A-Za-z0-9+/]\{4\}*([A-Za-z0-9+/]\{4\}
|[A-Za-z0-9+/]\{3\}=|[A-Za-z0-9+/]\{2\}==)})\\
 Subsequently, we manually scrutinized these payloads for any signs of malicious code, such as unusual system calls or requests to external servers. As depicted in Table~\ref{table:5}, the number of repositories containing malicious code is relatively low when compared to the total number of repositories employing base64 encoding. Nonetheless, we encountered numerous repositories that use base64 encoding to conceal their malicious proof of concept intentions. This underscores the fact that base64 encoding is one of the techniques employed to obfuscate malicious payloads.

\begin{table}[t]
\centering
\captionsetup{justification=centering}
\caption{Repositories with malicious base64 payloads }
\label{table:5}
\arrayrulecolor{black}
\begin{tabular}{ccc}
\hline
\textbf{Year}  &  \textbf{\# Repos} &  \textbf{ \# Malicious repos}   \\
\arrayrulecolor[rgb]{0.541,0.541,0.541}\hline
2017  &     6424        & 4         \\
\rowcolor[rgb]{0.804,0.804,0.804} 2018  &     7021        & 4        \\
2019  &     11336       & 16         \\
\rowcolor[rgb]{0.804,0.804,0.804} 2020  &     10588       & 28         \\
2021  &     14487        & 26 \\ \arrayrulecolor{black}\hline
Total  &     47285       & 74 (unique)     \\
\end{tabular}

\end{table}

\subsection{Summary of Results}\label{sec:summary}
Table~\ref{table:summary} shows the results regarding detection of maliciousness in PoCs for CVEs using the heuristics previously discussed. We detected in total 899 malicious repositories out of 47285. The number of malicious repositories is is considerably higher in years between 2019 and 2021 compared to 2017 and 2018. These years witnessed a few CVEs that have had a massive security impact. 

\begin{table}[t]
\centering
\captionsetup{justification=centering}
\caption{Summary of maliciousness detection}
\label{table:summary}
\begin{tabular}{ccc} 
\hline
\textbf{Year}                                   & \textbf{\# Repos} & \textbf{\# Malicious}  \\ 
\arrayrulecolor[rgb]{0.541,0.541,0.541}\hline
2017                                   & 6424     & 104           \\
\rowcolor[rgb]{0.804,0.804,0.804} 2018 & 7021     & 37           \\
2019                                   & 11336    & 272          \\
\rowcolor[rgb]{0.804,0.804,0.804} 2020 & 10588    & 243          \\
2021                                   & 14487    & 253          \\ 
\hline
Total                                  & 47285    & 899          \\
\end{tabular}

\end{table}

\subsection{A One-Year Review}\label{sec:longterm}

In May 2023, 13 months after the initial data collection process and 6 months after our disclosure to GitHub, we queried the same repositories to assess the current status of the gathered PoCs and their updates. This involved querying the information of each repository directly from GitHub. Our focus was twofold: first, we examined the availability of each repository, and second, we noted the date of the last update for each repository. Out of the initial 47,285 repositories collected, we discovered that 1,945 were no longer available. Notably, three of these repositories had been removed by GitHub staff due to violations of GitHub's terms of service. Two\footnote{\url{https://github.com/HarmvandenBrink/CVE-2021-36260}; Not available anymore at the time of writing.} \footnote{\url{https://github.com/matengfei000/cve-2019-0709}; Not available anymore at the time of writing.} of these removed repositories were also flagged as malicious by our analysis process. Additionally, we observed that 3,500 repositories had undergone changes within the 13-month period, either when a commit is pushed to any of the repository’s branches (551), or when the repository object undergone changes in the description or the primary language of the repository (3,334). The rest of repositories had remained unchanged (41,840). Note that a change to the repository object might be a commit. For a summary of these findings, we refer to Table~\ref{table:repos_by_mutations}. 

Among the discovered malicious PoC repositories, as of May 2023, 851 are still alive, and 48 were taken down.

\begin{table}[t]
\centering
\captionsetup{justification=centering}
\caption{State of PoC repositories after 13 months of collection}
\label{table:repos_by_mutations}
\begin{tabular}{cc} 
\hline
\textbf{Status} & \textbf{\# Repos} \\ \hline
Unchanged                 & 41840                  \\  \rowcolor[rgb]{0.804,0.804,0.804} 
Changed                 & 3500                  \\
Pushed               &   551               \\  \rowcolor[rgb]{0.804,0.804,0.804} 
Updated                 & 3334                  \\
Taken down by owner               & 1942                 \\  \rowcolor[rgb]{0.804,0.804,0.804} 
Taken down by GitHub                    & 3                \\ \hline
\end{tabular}

\end{table}

\section{Malicious PoCs Analysis}\label{sec:analysis}


\subsection{Meta Data Analysis}\label{sec:users}
\begin{figure}[t]
    \centering
    \captionsetup{justification=centering}
    \includegraphics[scale=0.38]{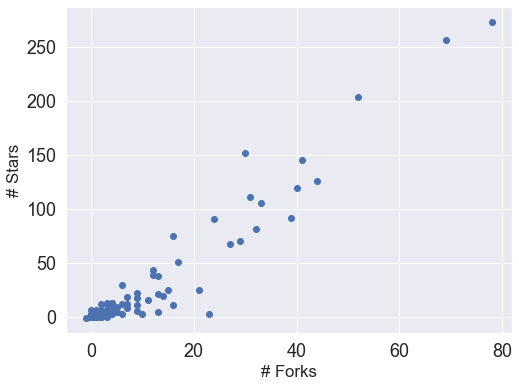}
  \caption{Distribution of malicious repositories based on stars and forks}
  \label{fig:Repos_Stars_Forks}
\end{figure}

GitHub repositories contain several fields that provide valuable information about the repository's activity and popularity~\cite{gonzalez2021anomalicious,wyss2022what}. Number of stars and forks are common metrics of popularity for GitHub repositories. As depicted in Figure~\ref{fig:Repos_Stars_Forks}, the majority of malicious repositories have less than 100 stars and 20 forks, although there are a few exceptions with higher counts. These outliers may be attributed to various factors, such as the repository's reputation. Overall, malicious PoC repositories are not very popular.

\begin{figure}[t]
    \centering
    \captionsetup{justification=centering}
    \includegraphics[scale=0.38]{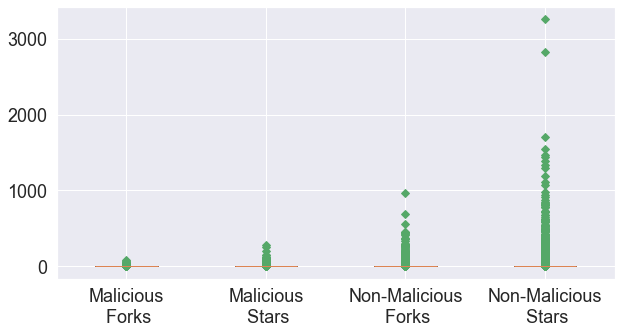}
    \caption{Boxplot diagrams of numbers of stars and forks for repositories that were discovered to be malicious and the remaining repositories (non-malicious).}
  \label{fig:forks_stars}
\end{figure}

Figure~\ref{fig:forks_stars} shows boxplots of the numbers of stars and forks for PoC repositories that we discovered to be malicious and the remaining repositories (non-malicious). We see that in both cases medians and the 3rd quartiles for both stars and forks are $0$, which means at least 75\% of repositories in both cases have no stars and no forks. Thus, stars and forks are not features that could be useful to identify malicious repositories, based on the data we observed.

In order to gain further insights into the malicious authors behind our dataset, we cross-referenced it with the HackerChatter\footnote{\url{http://www.hackerchatter.org/database}} dataset, which is an up to date, labelled malware GitHub repositories collected by Rokon et al.~\cite{rokon2020sourcefinder}. The results of this analysis provided valuable information about the presence of our dataset in other flagged malware source-code repositories.

Firstly, when considering our malicious data, we found that there were no matching repositories (identified by a tuple $\langle$repository name, owner username$\rangle$) in the HackerChatter dataset. This suggests that the specific malicious PoCs we identified have not been detected by the approach proposed in~\cite{rokon2020sourcefinder}. However, we identified 27 users from our malicious dataset who were mentioned in the HackerChatter dataset. This implies that these users have been associated with nefarious activities related to malware source code in Github.

On the other hand, when analyzing our complete dataset, which includes both malicious and non-malicious repositories, we discovered that there were 12 matching repositories in the HackerChatter dataset. This indicates that a portion of our dataset has been flagged as malware-related repositories, but in our case we don't define these PoCs as malicious. 

In comparison, within our complete dataset, there were 347 users mentioned in the HackerChatter dataset. This can be expected since we are dealing with a dataset of PoCs for exploits.

We counted the number of unique users in our data, and we found 18786 users responsible for our PoC repositories, 649 of them are owners of malicious repositories. Majority (61.95\%) of malicious repositories belong to users who own only one repository, while the highest number of malicious repositories per user is 50. Appendix~\ref{sec:users-appendix} presents more detailed information on distributions of repository ownership in our data.

\subsection{PoC Similarity Analysis}\label{sec:similarity}

In the previous section we have discussed how to identify malicious PoCs based on security-relevant indicators. We now aim to investigate whether code similarity can be useful in distinguishing between malicious and non-malicious PoCs. To assess this, we used JPlag\footnote{\url{https://github.com/jplag/JPlag}}, a code plagiarism detection system that helps detecting similarities between source code~\cite{prechelt2002finding}. It works by comparing source code files and identifying similarities between them to detect potential instances of code plagiarism or copying. JPlag was chosen as a plagiarism detection tool due to its effectiveness and widespread use~\cite{pauck2022scaling,ko2017coat,burrows2007efficient}.




The efficiency of JPlag lies in its approach to code comparison. It uses a combination of token-based and tree-based analysis techniques to detect similarities in the structure and logic of code~\cite{prechelt2002finding}. JPlag tokenizes the source code files and creates abstract syntax trees (ASTs) to represent the code's structure. It then performs similarity analysis by comparing the tokens and ASTs across different code files.

JPlag is efficient in detecting various types of code plagiarism, including verbatim copying, code restructuring, and code obfuscation~\cite{ko2017coat}. It employs a flexible threshold-based approach, allowing users to set similarity thresholds to customize the sensitivity of the detection  depending on their specific needs and requirements. One of the advantages of JPlag is its language support. It can analyze code written in multiple programming languages, including Java, C++, C\#, Go, Kotlin, Python, R, Rust, Scala, Scheme, EMF Metamodel\footnote{\url{https://github.com/jplag/JPlag\#supported-languages}}, and more. This versatility makes it suitable for detecting plagiarism in a wide range of PoCs. According to GitHub information, we have PoC repositories written in all these languages, except EMF Metamodel.

To measure similarity between different PoC repositories, we applied JPlag to all pairs of repositories written in the same JPlag-supported language that also target the same CVE IDs. We investigate similarity of original repositories among each other (removing forked ones). Due to the lack of space, we analyze similarity of original and forked repositories in Appendix~\ref{sec:similarity-appendix}.


\begin{figure}[t]
    \centering
    \captionsetup{justification=centering}
    \includegraphics[scale=0.32]{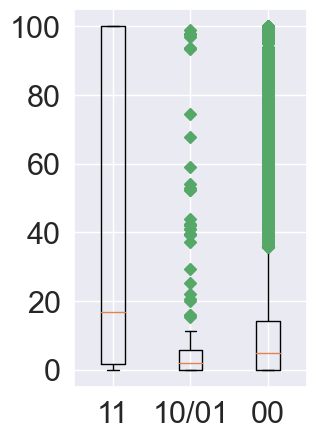}
    \caption{Similarity of original (non-forked) repositories among each other. We denote a malicious repository as \texttt{1} and a repository without a discovered malicious payload as \texttt{0}}
  \label{fig:original_boxplot}
\end{figure}

Figure~\ref{fig:original_boxplot} shows boxplots of similarity scores between all original repositories in our dataset (including pairs with 0\% similarity).  The similarity data in this figure is broken down into 3 categories:
\begin{itemize}
    \item Both compared repositories are malicious. We encode this case with label \texttt{11} (\texttt{1} stands for a malicious repository).
    \item One compared repository is malicious, and another one is not found to be malicious with our heuristics; encoded with label \texttt{10/01}.
    \item Both repositories are not found to be malicious with our approach; encoded with label \texttt{00}.
\end{itemize}

We can see that malicious repositories on average have slightly higher similarity than non-malicious repositories. This could be an indicator that code similarity is a useful feature to identify malicious repositories.

We performed the non-parametric Mann-Whitney U test to check a hypothesis that malicious repositories are more similar to each other than benign ones. The test rejected the null-hypothesis (that medians of similarity data for the \texttt{11} and \texttt{00} cases are equal) with $p$-value=0.034 ($<0.05$). The test also allows us to reject the null-hypothesis for the similarity data of the \texttt{11} and \texttt{01/10} cases (comparing malicious with each other versus comparing non-malicious repositories with malicious ones) with $p$-value=0.01 ($<0.05$). These results confirm that code similarity could be used for distinguishing malicious and benign PoCs.

\section{Case Studies}\label{sec:casestudies}

Throughout our research, we came across numerous instances of malicious PoCs. These proof of concepts served various purposes: some contained malware, others were used for gathering user information, and some were simply created to serve as reminders, mocking individuals who run proof of concepts without reading the accompanying code and recognizing the potential harm involved. In this section and Appendix~\ref{sec:morecasestudies}, we will delve into each of these types and provide an example for better understanding.

\paragraph{Houdini Malware.}
An intriguing example we encountered during our research was a repository that aimed to serve as a proof of concept for \texttt{CVE-2019-0708}\footnote{\url{https://github.com/Elkhazrajy/CVE-2019-0708-exploit-RCE}; Not available anymore at the time of writing.}, famously known as BlueKeep\footnote{\url{https://www.cve.org/CVERecord?id=CVE-2019-0708}}. This repository was created by username \texttt{Elkhazrajy}.
  
Upon inspecting the source code, we identified a base64-encoded line, which, once decoded, initiated the execution of another Python script. This script contained a link to \texttt{Pastebin}\footnote{\url{https://pastebin.com}}, where a VBScript was hosted. The first \texttt{exec} command in the Python script executed this VBScript. Further investigation of the VBScript revealed the presence of the Houdini malware. We provide the screenshots of our analysis in Appendix~\ref{sec:morecasestudies}. Houdini, also known as H-Worm, has been active since at least 2013\footnote{\url{https://www.mandiant.com/resources/blog/now-you-see-me-h-worm-by-houdini}} and still being active today\footnote{\url{https://www.trellix.com/en-us/about/newsroom/stories/research/cyberattacks-targeting-ukraine-increase.html}}. Houdini malware is typically distributed through various means, such as malicious email attachments, exploit kits, or social engineering techniques.  

We discovered similar techniques employed in other repositories, where malware samples were double obfuscated. These samples established communication with external hosts, downloaded malicious files, and subsequently executed them using VBScript. These malware samples predominantly targeted Windows systems. The host used in the Houdini code mentioned above was not functional at the time of writing.

\paragraph{CobaltStrike.} 
Another instance involving malicious binaries can be observed in a repository that was previously accessible at \footnote{\url{https://github.com/matengfei000/cve-2019-0709}; Not available anymore at the time of writing.}. This repository contained a binary file purportedly serving as an exploit for \texttt{CVE-2019-0709}. Notably, the binary was flagged by VirusTotal as a CobaltStrike instance\footnote{\url{https://www.virustotal.com/gui/file/fccc5846bd9e09b8f05d4628b684bc4d3ee105280d8ad8c8607a3c6fe746bbaa/detection}}. CobaltStrike provides a C\&C solution frequently used by adversaries. 
Although the repository is no longer accessible, the user behind it still exists under a different username, as indicated by a GitHub redirect. According to VirusTotal, this binary was initially scanned on March 22, 2018, and was last submitted on May 15, 2019. Interestingly, this submission date is approximately one month before the release of CVE details in the NVD database\footnote{\url{https://nvd.nist.gov/vuln/detail/CVE-2019-0709}}.

\paragraph{RokRAT.}
Is a type of malware that belongs to the RAT family and was linked back to APT37\footnote{\url{https://malpedia.caad.fkie.fraunhofer.de/actor/apt37}}. RokRAT is designed to infiltrate systems, collect sensitive information, and enable the attacker to perform various malicious activities, such as stealing credentials, logging keystrokes, capturing screenshots, and executing arbitrary commands. It can also establish a backdoor on the compromised system, providing persistent access for the attacker. RokRAT was introduced back in 2017\footnote{\url{https://blog.talosintelligence.com/introducing-rokrat/}} and still being used until today\footnote{\url{https://research.checkpoint.com/2023/chain-reaction-rokrats-missing-link/}}. During our investigation, we found six binaries that were flagged by VirusTotal as RokRAT\footnote{\url{https://www.virustotal.com/gui/file/e1546323dc746ed2f7a5c973dcecc79b014b68bdd8a6230239283b4f775f4bbd/}}. These six binaries were found to be related to the same \texttt{CVE-2017-4878}\footnote{\url{https://nvd.nist.gov/vuln/detail/CVE-2017-4878}}, which appears to be rejected by NVD.

Due to the page limit, we discuss more case studies in Appendix~\ref{sec:morecasestudies}.


\section{Discussion}\label{sec:discussion}

Using the proposed heuristics we have been able to detect that at least 1.9\% of repositories in our data contain malicious PoCs. This rate of untrustworthy exploits, although relatively low, is still a cause for concern, as they are being used by security practitioners across the world. We thus believe that our study is an important first step towards creating more reliable tools for security professionals.

We note that GitHub's acceptable use policy\footnote{\url{https://docs.github.com/en/site-policy/acceptable-use-policies/github-active-malware-or-exploits}} allows publication of security-relevant content, such as malware, vulnerability information of exploits, but only for research purposes. They explicitly prohibit dual-use exploits (so, i.a., malicious exploits) and they require all exploit data to be clearly labelled as potentially harmful content. We have not observed any warnings in the subset of repositories that we inspected manually. Moreover, the discovered 899 repositories with malicious PoCs are clearly in violation of GitHub's policy. Our findings have been reported to GitHub in October 2022, and, as we mentioned, since then 3 of these repositories have been taken down by GitHub, although we do not know whether it was done because of our reporting or not. GitHub has not reached back to us since our reporting. 

\paragraph{Limitations.}
Our study has several limitations. First, the GitHub API proved unreliable and not all repositories corresponding to the used CVE IDs were collected. Therefore, we potentially have only a subset of data related to the targeted CVEs. We have tried to mitigate this issue by re-querying the API a second time.

Another limitation of this study is that we rely on heuristics for detecting malicious PoCs. These heuristics are likely not able to discover all malicious PoCS in our dataset, as some of them could apply more substantial obfuscation techniques, such as encryption. At the same time, due to the amount of time that passed since older exploits were published, it is possible that some previously malicious IP addresses are not detected as malicious anymore. Thus, we might have missed malicious PoCs that include such ``previously malicious'' IPs.

These limitations could potentially be addressed in the future by developing more sophisticated dynamic analysis techniques that would execute PoCs and observe their behavior. However, given the variety of technologies the PoCs are developed with (see Table~\ref{table:prog_lang}), the variety of PoC designs (arguments handling, etc.), and the variety of the targeted platforms, it is very challenging to develop a single platform to analyze a significant proportion of the PoCs in our dataset. We also think that static analysis can be advantageous over dynamic analysis when investigating code maliciousness \emph{in the past}, particularly in scenarios where the code interacts with external malicious websites that are no longer accessible but were previously flagged by third-party services like VirusTotal. This is why in the proposed method we relied on indicators that can be extracted with static analysis.

Another viable alternative is to develop a detection mechanism based on code similarity analysis that will leverage machine learning techniques, similar to, e.g., the Amalfi approach to detect malicious npm packages~\cite{sejfia2022practical}. Our results show that code similarity can indeed be a reliable indicator to discriminate malicious PoCs from non-malicious ones.

Still, as with any complex security-related task, such as e.g., malware detection, adversaries are able to apply very sophisticated obfuscation mechanisms to hide the malicious PoC functionality from inspection, therefore a 100\% reliable detection mechanism for malicious PoCs is not achievable.

\paragraph{Ethical considerations.}
Ensuring ethical practices and responsible handling of potentially malicious code is of paramount importance in our research. To mitigate any potential risks and safeguard against unintended consequences, we followed a few rules when executing some selected PoCs obtained from GitHub to better understand their behavior (reported as case studies).

First, a careful manual review process was conducted to evaluate the nature and intent of the PoCs. This involved scrutinizing the code for any indications of malicious behavior, such as direct calls to known malicious IPs or suspicious payloads. This review was conducted by one of the authors who is an experienced security analyst and has extensive knowledge about malware analysis and reverse engineering. Then, by running the PoCs in a sandboxed environment, we were able to monitor their behavior, assess their impact, and analyze any potential malicious activities without endangering the integrity of our infrastructure. After careful examination of the code and the behavior in sandbox, the analyst took decisions how to further examine the case study PoCs (e.g., about retrieving new PoC components from remote websites). This approach allowed us to strike a balance between conducting meaningful research on PoCs' functionality, while minimizing any potential harm to systems or networks.  

\section{Related Work}\label{sec:relwork}

To the best of our knowledge, the problem of malicious exploit PoCs in GitHub has not yet been studied in the literature. But the community has addressed several important related issues.

\paragraph{Detecting malicious third-party code.}
Malicious third-party code detection is a problem that has been investigated in context of software supply chains, where large ecosystems have been found to be infected with malicious packages. For example, researchers investigated approaches to identify malicious packages in the npm, PyPI and Java ecosystems (e.g.,~\cite{scalco2022feasibility,vu2022benchmark,sejfia2022practical,ladisa2022towards,ohm2022feasibility}). Some of the heuristics we utilize for detecting malicious PoCs (IP addresses, obfuscation detection, malicious binaries) have been applied in those studies in the context of detecting malicious software packages. In our case, the challenge is to distinguish an unexpected adversarial functionality in the context of PoCs, which are by definition considered malicious if applied by black hat hackers.


In the area of malware detection, the existing literature focuses on capturing semantic representations of code behavior (e.g., sensitive API calls~\cite{ladisa2022towards,sejfia2022practical,tam2017evolution}). We note that the heuristics we propose in this work are more lightweight, because we deal with a very diverse code base (many different programming languages and different underlying systems to be exploited by PoCs).

\paragraph{Detecting malicious code in GitHub.}
The community has also studied peculiarities of detecting malicious activity on GitHub. 
For example, Gonzales et al.~\cite{gonzalez2021anomalicious}  analyzed the user's profile information and commit logs to identify anomalous commits that represent potential malicious activities in a GitHub repository.  Qian et al.\cite{qian2022malicious} developed a framework called Heterogeneous Graph to detect malicious repositories by leveraging relationships and meta data.



Code similarity is a widely used technique to cluster malicious snippets and detect closely related ones~\cite{zhang2021survey,li2019rebooting}. In the context of GutHub, Cao and Dolan-Gavitt~\cite{Cao2022WhatTF} proposed a solution to detect malicious forks on GitHub by checking known malicious signatures and computing included file similarity using the \texttt{ssdeep} algorithm. 

We have shown that code similarity as measured by JPlag~\cite{prechelt2002finding} can be used to cluster malicious and benign PoCs. In the future, this approach can be further improved by utilizing techniques developed for detecting similar GitHub projects, e.g.,~\cite{nguyen2020automated,rokon2021repo2vec,qian2022rep2vec}, and identifying hidden~\cite{wyss2022what} or malicious~\cite{Cao2022WhatTF} forks.

Rokon et al.~\cite{rokon2020sourcefinder} have identified 7.5 thousands of malware source code repositories on GitHub. They have used manual analysis to label 1000 repositories, and then they have built a classifier using words from repository title and description. We report on our cross-check with their dataset in Section~\ref{sec:users}.


\paragraph{CVE exploits.}
%
Researchers have studied the life cycle and exploitation likelihood of CVEs using various methodologies~\cite{le2022survey}. Analysis of social network data has been very prominent in this space. Horawalavithana et al.\cite{horawalavithana2019mentions} examined the discussions surrounding new CVEs on platforms like GitHub, Reddit, and Twitter. Sabottke, Suciu, and Dumitras~\cite{sabottke2015vulnerability} focused on detecting information about CVE exploitation being shared on Twitter. They developed methods to identify relevant discussions and extract valuable insights regarding CVE exploitation from the platform. 

Schiappa, Chantry and Garibay~\cite{schiappa2019cyber} have investigated social media discourse concerning CVEs and exploits on Twitter, GitHub and Reddit. In their data, in 2015 and 2016 on average 15\% of CVEs were associated with PoCs available in ExploitDB. This is much higher than what we observed in our data (for 2017 onwards). Study~\cite{schiappa2019cyber} has shown GitHub to be a useful source of information concerning possible attack vectors. Shrestha et al.~\cite{shrestha2020multiple} have reached similar conclusions, as they found that CVE-related information is being disseminated in GitHub discussions even prior to a vulnerability being officially published. Neil, Mittal and Jishi~\cite{neil2018mining} have further demonstrated feasibility of automatically mining vulnerability-related threat intelligence from GitHub and similar version control platforms.

Suciu et al.\cite{suciu2022expected} focused on measuring the likelihood of exploits becoming functional over time for a given CVE. Yang et al.~\cite{yang2020better} used machine learning algorithms to predict the likelihood of a CVE being exploited based on the existence and source of a PoC. Householder et al.\cite{householder2020historical} investigated the development of CVE exploits over time and analyzed the chances of a CVE having an exploit in the future. They provided statistics, indicating the percentage of CVEs with publicly available exploits, which aligns with our findings (Table~\ref{table:CVEs_unique}).


Al Alsadi et al.~\cite{al2022no} studied IoT exploits extracted from malware binaries using static and dynamic analysis techniques. Like us, the study~\cite{al2022no} applied VirusTotal to classify malicious binaries. They observed packing and obfuscation as common techniques used by malware for hindering detection. Yet, the exploit use case investigated by in~\cite{al2022no} is different from ours: their study aimed to dissect exploits hidden in IoT binaries, while we focus on PoCs that seemingly offer exploitation capabilities for known CVEs to security researchers. 

These studies collectively contribute to the understanding of different aspects related to malicious code, repositories, commits, forks, and CVE exploits on platforms like GitHub and Twitter. Although these studies provide valuable insights, none of them specifically focused on analyzing repositories containing PoCs for CVE exploits and assessing their reliability. Therefore, our research stands as the first of its kind in this particular area.


\section{Conclusions}\label{sec:conclusions}

We conducted an extensive investigation into the maliciousness of CVE Proof of Concepts (PoCs) on GitHub.  Out of the 47,285 GitHub repositories containing PoCs, we detected 899 malicious repositories, which accounts for approximately 1.9\% of the total. 

To the best of our knowledge, our work represents the first comprehensive investigation that analyzes PoCs of CVEs hosted on public platforms such as GitHub and proposes methods for detecting malicious PoCs. Our approach involves examining the source code for indicators of malicious activity, such as calls to suspicious servers, extracting hexadecimal payloads, and identifying base64-encoded scripts that may contain malicious instructions. These instructions could include actions such as exfiltrating information, downloading malicious files, or creating backdoor access. However, we note that our current approach may not detect every malicious PoC, as there will always be new creative ways to obfuscate code. In the light of this, we have explored the use of code similarity as a feature to aid in identifying new malicious repositories. Our findings indicate that, on average, malicious repositories exhibit higher levels of similarity to each other compared to non-malicious ones. This result is an important step towards the development of more robust detection techniques, what will be our goal in the future work.

Our chosen approach emerged after examining numerous PoCs and observing that the majority employed common techniques, such as hexadecimal and base64 payloads, which may not immediately raise suspicion among PoC users. We believe that our findings make a significant contribution for the security community. Within these malicious PoCs, we discovered instructions aimed at opening backdoors or planting malware on the systems running them. Consequently, these PoCs specifically target the security service community, posing a risk to every customer of such a security company utilizing PoCs sourced from GitHub. By identifying these malicious PoCs, we strive to raise awareness of risks associated with specific repositories and assist the security community in exercizing caution when engaging with them.



\bibliographystyle{plain}
\bibliography{FakePoCs}

\newpage
\appendix

\section{Repository Ownership Information}\label{sec:users-appendix}

Repository ownership break-down in our data is presented in Table~\ref{table:repos_users_all}. Table~\ref{table:repos_users} presents the distribution of ownership of malicious repositories. 

\begin{table}[h]
\centering
\captionsetup{justification=centering}
\caption{Distribution of malicious repository ownership}
\label{table:repos_users}
\begin{tabular}{cc}
\hline
\textbf{\# Malicous repos per user}  &  \textbf{Count}    \\
\arrayrulecolor[rgb]{0.541,0.541,0.541}\hline
50  &     1 \\
\rowcolor[rgb]{0.804,0.804,0.804} 20  &     1 \\
15  &     1  \\
\rowcolor[rgb]{0.804,0.804,0.804} 14  &     1\\
9  &     1  \\
\rowcolor[rgb]{0.804,0.804,0.804} 8  &     3\\
7  &     2 \\
\rowcolor[rgb]{0.804,0.804,0.804} 6  &     1\\
5  &     1 \\
\rowcolor[rgb]{0.804,0.804,0.804} 4  &     4\\
3  &     17 \\
\rowcolor[rgb]{0.804,0.804,0.804} 2  &     59\\
1  &     557  \\ \arrayrulecolor{black}\hline
\end{tabular}
\end{table}

\begin{table}[h]
\centering
\captionsetup{justification=centering}
\caption{Distribution of (total) repository ownership}
\label{table:repos_users_all}
\begin{tabular}{cc}
\hline
\textbf{\# Repos per user} & \textbf{Count} \\ \hline
1459              & 1     \\ \hline
\rowcolor[rgb]{0.804,0.804,0.804}
401-600           & 2     \\ \hline
201-300           & 3     \\ \hline
\rowcolor[rgb]{0.804,0.804,0.804}
101-200           & 13    \\ \hline
91-100            & 2     \\ \hline
\rowcolor[rgb]{0.804,0.804,0.804}
81-90             & 6     \\ \hline
71-80             & 5     \\ \hline
\rowcolor[rgb]{0.804,0.804,0.804}
61-70             & 6     \\ \hline
51-60             & 11    \\ \hline
\rowcolor[rgb]{0.804,0.804,0.804}
41-50             & 38    \\ \hline
31-40             & 39    \\ \hline
\rowcolor[rgb]{0.804,0.804,0.804}
21-30             & 95    \\ \hline
0-20              & 18565 \\ \hline
\end{tabular}
\end{table}

\section{Similarity of Original and Forked PoCs}\label{sec:similarity-appendix}.

\begin{figure}[t]
    \centering
    \captionsetup{justification=centering}
    \includegraphics[scale=0.32]{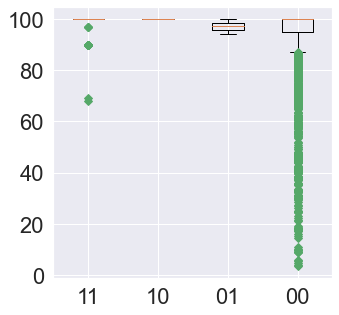}
    \caption{Similarity of forked repositories as computed by JPlag (only pairs with positive similarity score). We denote a malicious repository as \texttt{1} and a repository without a discovered malicious payload as \texttt{0}}
  \label{fig:forked_boxplot}
\end{figure}


We now analyze similarity of the original and forked repositories.  Figure~\ref{fig:forked_boxplot} shows the similarity data between the original and forked repositories (we only visualize pairs that have a positive similarity score according to JPlag). The similarity data in this figure is broken down into 4 categories:
\begin{itemize}
    \item Both original and forked repositories are malicious. We encode this case with label \texttt{11} (\texttt{1} stands for a malicious repository).
    \item Parent repository is malicious, but the forked repository is not found to be malicious with our heuristics. We encode this case with label \texttt{10}.
    \item Parent repository is not found to be malicious, but the forked repository is malicious; encoded with label \texttt{01}.
    \item Both repositories are not found to be malicious with our approach; encoded with label \texttt{00}.
\end{itemize}

We can see in Figure~\ref{fig:forked_boxplot} that all cases (\texttt{11}, \texttt{10}, \texttt{01} and \texttt{00}) have high similarity (median similarity is 100\% or close to it). By analyzing similarities between original repositories and forked ones we have 3 interesting cases of similarities which we investigated manually, which are the following:
\begin{enumerate}
  \item Original repository is malicious and the forked repository is malicious, but their similarity is less than 100\%. In this case, it is interesting to see what modifications have been done to the malicious PoC. We have 17 such pairs.
  \item Original repository is malicious, but the forked repository is not malicious (the cases labelled with \texttt{10}). In this case, it is interesting to see why the repository is now not detected to be malicious: whether the reason is that our heuristics are not reliable, or whether the user has changed something in the re-distributed PoC code and made it not malicious. We have 6 such pairs.
  \item Original repository is not malicious, but the forked repository is malicious (the cases labelled with \texttt{01}). We have 2 such pairs.
\end{enumerate}

We have manually checked the repositories from these cases and concluded that in of all these 3 scenarios the changes (in the similarity score or the malicious status) occurred because of inclusion/exclusion of a malicious EXE binary. No other types of modifications were discovered. This shows that inclusion of a malicious binary is a strong indicator of maliciousness.

One interesting case that we observed was a repository originally shared by a prominent European security company. This repository was forked, and this pair corresponds to the \texttt{01} case in our classification (parent is not malicious, but the forked repository is malicious). However, when we looked at the history of the original repository updates, we discovered that a malicious EXE file was originally included there. It was later deleted by the security company, but it remained in the forked repository. This case shows that even very experienced security professionals fall for malicious PoCs and can redistribute them.



\section{More Case Studies}\label{sec:morecasestudies}

\paragraph{Screenshots for the Houdini malware case study.}

\begin{figure*}[t!]
    \centering
    \captionsetup{justification=centering}
    \includegraphics[scale=0.555]{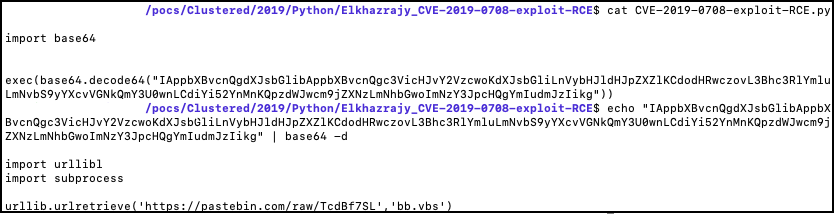}
  \caption{Houdini malware found in \texttt{CVE-2019-0708-exploit-RCE} repository by user \texttt{Elkhazrajy}}
  \label{fig:Malware_Example}
\end{figure*}

\begin{figure*}[t!]
    \centering
    \captionsetup{justification=centering}
    \frame{\includegraphics[scale=0.4]{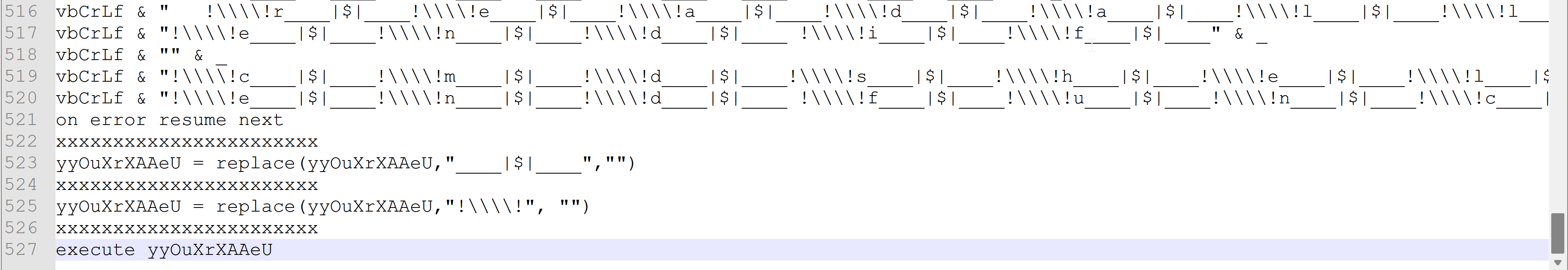}}
    \caption{An obfuscated payload in a PoC \texttt{CVE-2019-0708}}
  \label{fig:houdini1}
\end{figure*}

\begin{figure*}[t!]
    \centering
    \captionsetup{justification=centering}
    \frame{\includegraphics[scale=0.4]{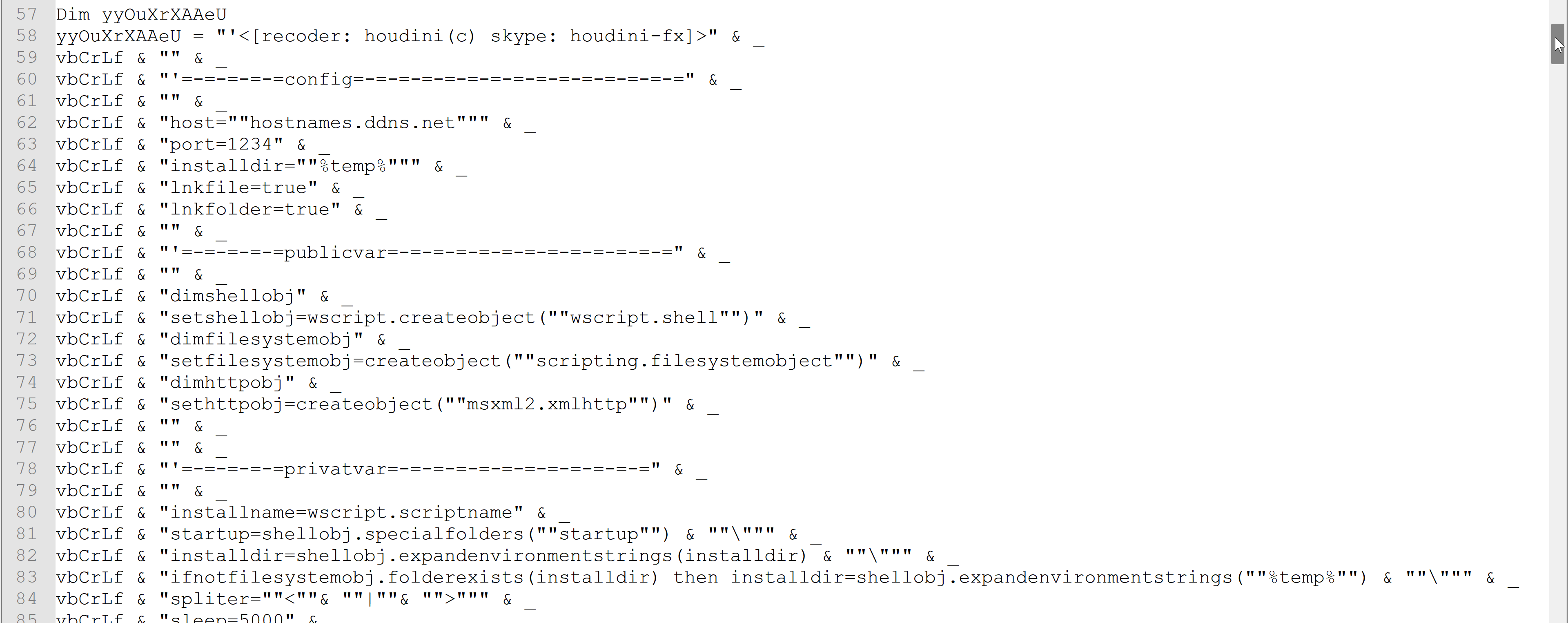}}
    \caption{Houdini malware after de-obfuscation}
  \label{fig:houdini2}
\end{figure*}

Screenshot of the Houdini malware functionality in the analyzed VBScript is shown in Figure~\ref{fig:Malware_Example}.
Screenshots in Figure~\ref{fig:houdini1} and Figure~\ref{fig:houdini2} showcase the VBScript code before and after de-obfuscation, respectively.

\paragraph{Exfiltration scripts.}
The purpose of the script in question is to exfiltrate telemetry data from the server, including the commands executed by the user. It also generates fake output to create the illusion that the attack has been successful. Listing~\ref{lst:fake_exfiltrate} provides a visual representation of the script. The base64 payload within the script contains a URL pointing to the server where the data is being exported.

We discovered this script in several repositories, either in the form of forks or copies with slight modifications, primarily involving the base64 encoded IP address.
  
\lstset{style=mystyle}
\begin{lstlisting}[language=Python, caption=Malicious PoC Exfiltration Example, label={lst:fake_exfiltrate}]
time.sleep(3)
lhost = os.uname()[1]
command = getpass.getuser() + '@' + (lhost)
args = ' '.join(sys.argv[1:])
ErrorMsg = 'Connection Terminated: (Timeout)'
URL = base64.b64decode('aHR0cDovLzU0LjE4NC4yMC42OS9wb2MyLnBocA==')
PARAMS = {'host':command, 'args':args, 'cve':Bug}
r = requests.get(url = URL, params = PARAMS)
welcome = r.content
if welcome != "":
	rsp = 1
	while rsp != "":
		cmd = raw_input(welcome)
		PARAMS = {'host':command, 'args':cmd, 'cve':Bug}
		r = requests.get(url = URL, params = PARAMS)
		rsp = r.content
		print rsp
		welcome = "C:\WINDOWS\system32>"
time.sleep(10)
\end{lstlisting}

\begin{figure}[h!]
  \centering
  \begin{minipage}[b]{0.48\textwidth}
    \frame{\includegraphics[width=\textwidth]{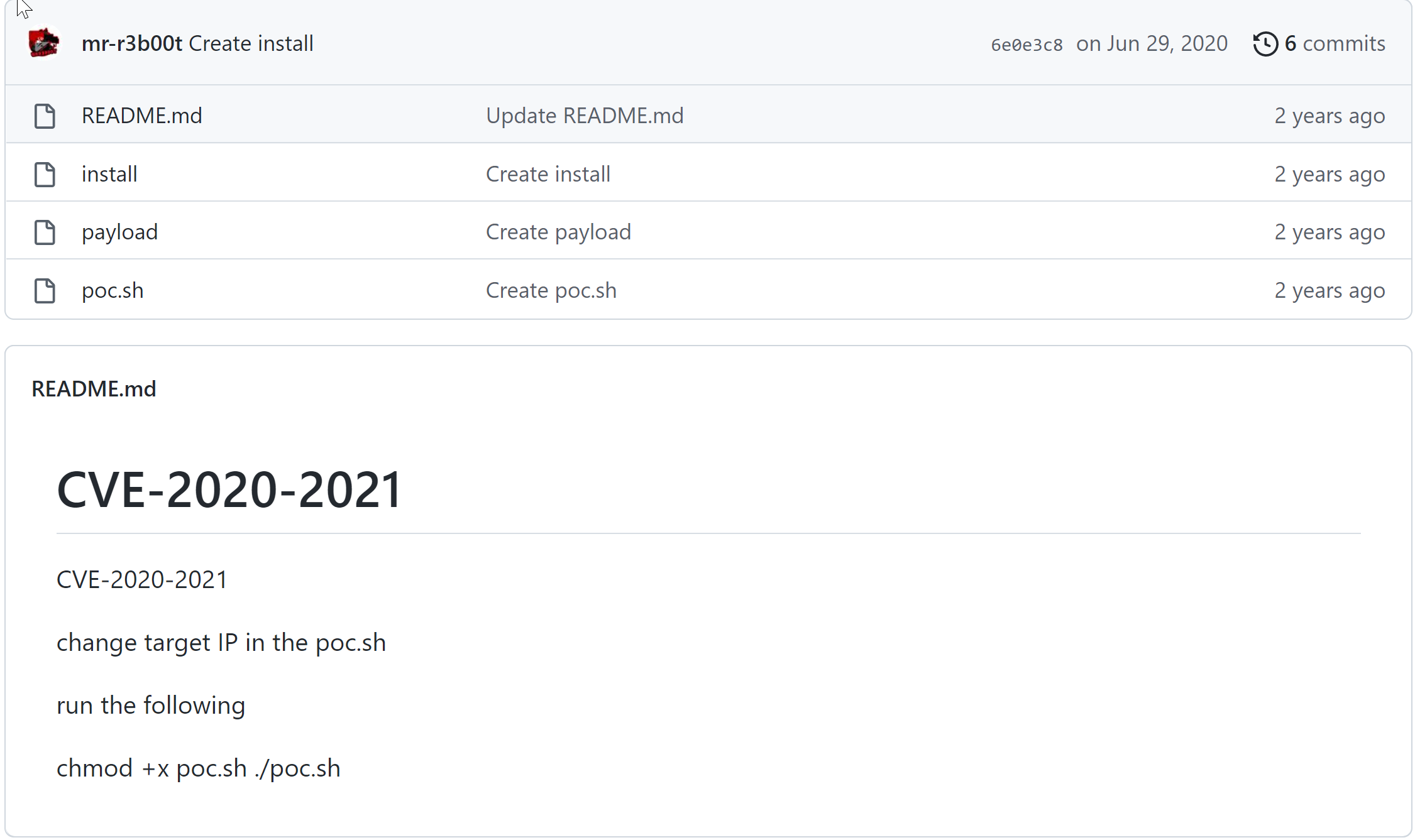}}
    \caption{Repository with fake PoC as a prank}
    \label{fig:prank1}
  \end{minipage}
  \hfill
  \begin{minipage}[h!]{0.4\textwidth}
    \includegraphics[width=\textwidth]{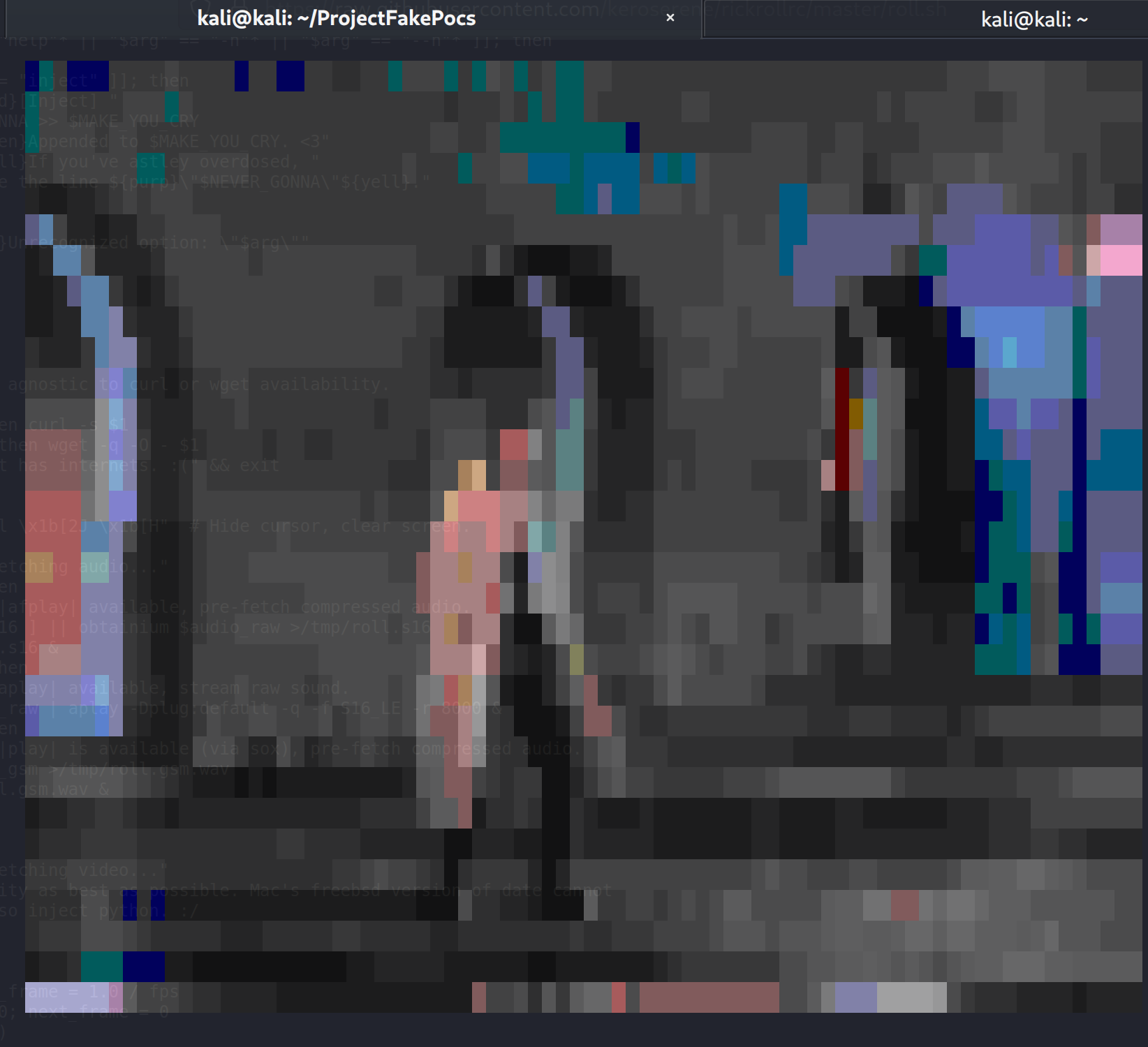}
    \captionsetup{justification=centering}
    \caption{Rickroll prank in a fake PoC for \texttt{CVE-2020-2021}}
    \label{fig:prank2}
  \end{minipage}
\end{figure}

\paragraph{Prank scripts.}
These scripts are not malicious, but rather created by individuals who are aware of the security issue and aim to educate the community by sharing prank scripts. When executed, these scripts typically display prank messages or perform other harmless actions.
For example, in one of the repositories focused on the \texttt{CVE-2020-2021} (an Authentication bypass in PAN-OS\footnote{\url{https://docs.paloaltonetworks.com/pan-os}}), there is a Shell script designed to serve as the proof-of-concept (PoC). When executed, this script amusingly transforms the user's screen into a ``Rickroll'' prank, a popular internet meme involving the unexpected redirection to the music video for Rick Astley's ``Never Gonna Give You Up'', Figure~\ref{fig:prank1} and ~\ref{fig:prank2}.

\end{document}